\PassOptionsToPackage{prologue,dvipsnames}{xcolor}
\documentclass[aps,prl,twocolumn,showpacs,groupedaddress,superscriptaddress,footinbib,longbibliography]{revtex4-2}
\usepackage{natbib}
\usepackage{mathrsfs}
\usepackage{hyperref}
\setcitestyle{square,sort&compress,comma,numbers}
\hypersetup{colorlinks=true, citecolor=ForestGreen, urlcolor=ForestGreen, linkcolor=blue}
\usepackage[english]{babel}
\usepackage[utf8]{inputenc}
\usepackage{graphicx}
\usepackage{amsmath}
\usepackage{amsfonts}
\usepackage{amssymb}
\usepackage{color,soul}
\setulcolor{red}
\usepackage{easyReview}
\usepackage{xcolor}

\usepackage{balance}

\usepackage[normalem]{ulem}

\begin{document}
\title{Active Brownian Dynamics in Channels: First-Passage and Spatiotemporal Properties via Siegmund Duality}
\author{Yanis Baouche}
\email{contributed equally}
\affiliation{Max-Planck-Institut f{\"u}r Physik komplexer Systeme, N{\"o}thnitzer Stra{\ss}e 38,
01187 Dresden, Germany}
\author{Mathis Guéneau}
\email{contributed equally}
\affiliation{Max-Planck-Institut f{\"u}r Physik komplexer Systeme, N{\"o}thnitzer Stra{\ss}e 38,
01187 Dresden, Germany}
\author{Christina Kurzthaler}
\email{ckurzthaler@pks.mpg.de}
\affiliation{Max-Planck-Institut f{\"u}r Physik komplexer Systeme, N{\"o}thnitzer Stra{\ss}e 38,
01187 Dresden, Germany}
\affiliation{Center for Systems Biology Dresden, Pfotenhauerstra{\ss}e 108, 01307 Dresden, Germany}
\affiliation{Cluster of Excellence, Physics of Life, TU Dresden, Arnoldstra{\ss}e 18, 01062 Dresden, Germany}


\begin{abstract} 
Accumulation at boundaries represents a widely observed phenomenon in active systems with implications for microbial ecology and engineering applications. To rationalize the underlying physics, we study the first-passage properties and spatial distributions of an active Brownian particle (ABP) in a channel. Leveraging Siegmund duality, we establish a direct mapping between the propagators of ABPs with absorbing and hard-wall boundary conditions, yielding analytical results in both problems. We analyze the system across low and high activity regimes -- quantifying persistent motion relative to diffusion -- and show that active motion, together with a favorable initial orientation, typically lowers the mean first-passage time relative to passive diffusion.  Notably, the full time-dependent propagator between hard walls approaches a wall-accumulated stationary state, given by the derivative of the splitting probability as a consequence of Siegmund duality.
\end{abstract}

\maketitle

 \begin{figure}
     \centering
     \includegraphics[width=0.45\textwidth]{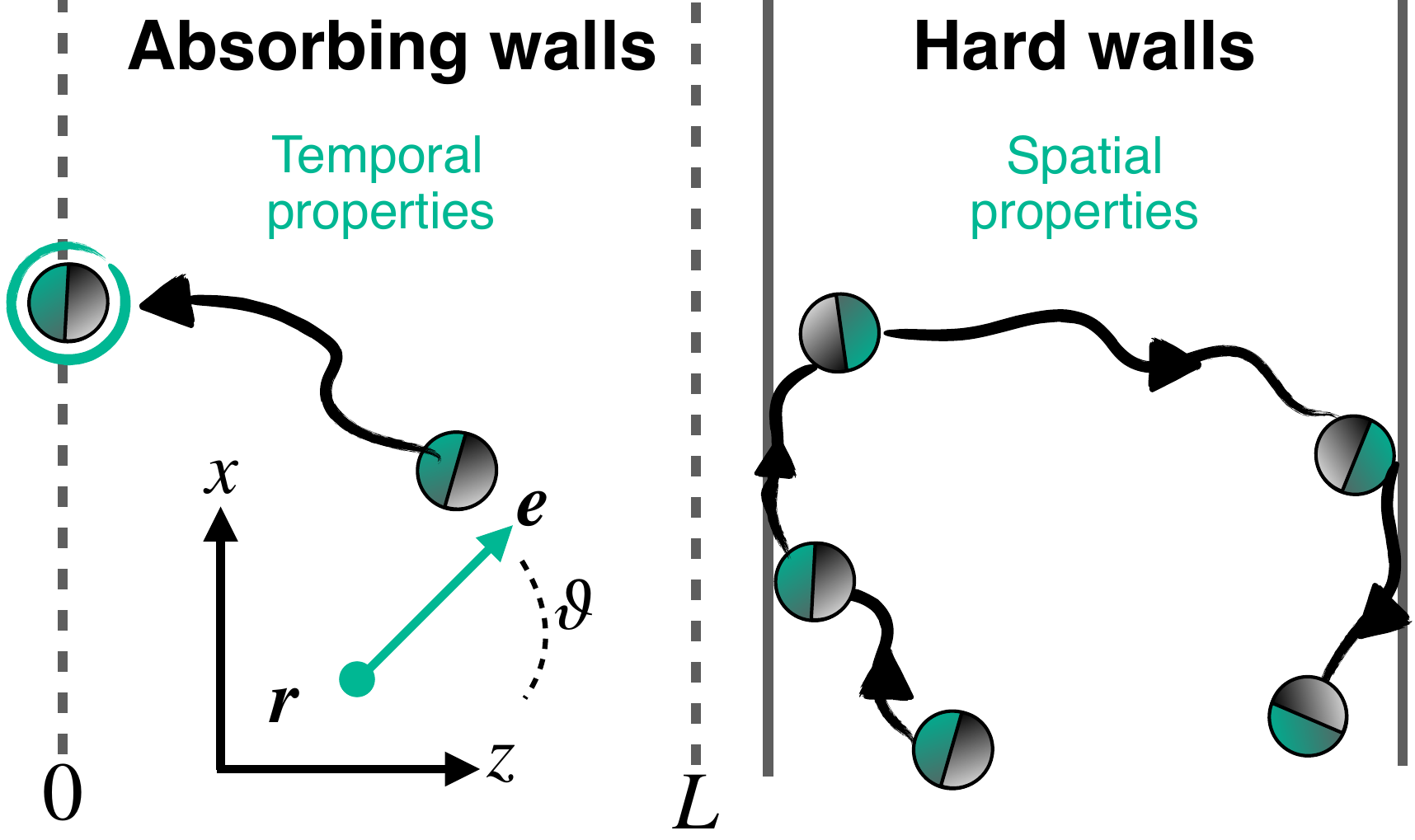}
     \caption{Schematic of the duals: ABPs between absorbing (sticking) and hard (reflective) walls. The left side shows an ABP hitting an absorbing boundary, which gives access to first-passage properties. The right side illustrates the dynamics between hard walls, yielding particle distributions. The inset depicts the agent's position $\boldsymbol{r}$ and orientation~$\vartheta$.}
     \label{fig:1_schematic}
 \end{figure}
Active agents inevitably engage with boundaries: not only do the latter facilitate foraging and survival of microorganisms via surface-concentrated nutrients and biofilm formation, they also mediate their navigation by enabling persistent exploration 
in the absence of complex sensing behavior~\cite{DiLeonardo:2010, Ray:2014, Galajda:2007, Bechinger:2016, Simmchen:2016, Wong:2021, Hallatschek:2023, Denissenko:2012}. 
Boundaries are also the pivot point for functional tasks, where future microrobots are expected to translate their energy into mechanical work for performing therapeutic or engineering operations~\cite{gompper2025MotileActive2025, Baulin:2025, Volpe:2025, erkocMobileMicrorobotsActive2019, alapanMicroroboticsMicroorganismsBiohybrid2019,nauberMedicalMicrorobotsReproductive2023}. To guide the design of these applications and obtain a physical understanding of the microbiological phenomena, it is crucial to unravel two questions: first, how long does it take an active agent to reach a boundary and, second, how do boundaries alter their distributions. 

To address these topics, a number of models -- including the canonical active Brownian particle (ABP)~\cite{ditrapaniActiveBrownianParticles2023,baoucheFirstpassagetimeStatisticsActive2025, iyaniwuraMeanFirstPassage2025,basuActiveBrownianMotion2018, basuLongtimePositionDistribution2019, poncet_pair_2021, caraglio_two-dimensional_2026, santra_active_2021, moen_trapping_2022}, the active Ornstein–Uhlenbeck particle~\cite{woillezNonlocalStationaryProbability2020}, and the run-and-tumble particle~\cite{angelaniFirstpassageTimeRunandtumble2014,dharRunandtumbleParticleOnedimensional2019,malakarSteadyStateRelaxation2018,gueneauOptimalMeanFirstpassage2024,gueneauRunandtumbleParticleOnedimensional2025,debruyneSurvivalProbabilityRunandtumble2021,moriUniversalSurvivalProbability2020} -- have shown how the intrinsic dynamics leverage activity to enable accelerated target reaching in comparison to passive particles.
Confined dynamics also display characteristic phenomena: highly-active particles can accumulate near walls~\cite{elgetiSelfpropelledRodsSurfaces2009,elgetiWallAccumulationSelfpropelled2013} without the need of hydrodynamics, while the time it takes agents to escape a wide, slowly varying channel has been linked to their distribution near the boundaries~\cite{zhaoActiveParticlesTube2025}, reflecting the curvature-dependent pressure field of active systems~\cite{solonPressureNotState2015,smallenburgSwimPressureWalls2015,duzgunActiveBrownianParticles2018}. The phenomenology of boundary interactions is thus rich, yet the coupling between translational and rotational degrees of freedom makes deriving analytical expressions for confined active dynamics challenging and requires new theoretical frameworks. 

A particularly useful tool is Siegmund duality~\cite{siegmundEquivalenceAbsorbingReflecting1976, gueneauSiegmundDualityPhysicists2024}, which establishes an exact correspondence between stochastic dynamics with absorbing and hard boundaries (Fig.~\ref{fig:1_schematic}).
Specifically, for two walls separated by $L$, the hard-wall propagator $p_{H}(z,t|z_0)$ (i.e., the probability density to be at position $z$ at time $t$, given initial position $z_0$), follows from the absorbing-wall propagator~$p_{A}$~via:
\begin{equation}
\label{eq:propagator_relation}
 p_H(z,t \, |\,   z_{0})
=
\int_{z_0}^{L}\!dz'\;\frac{\partial}{\partial z}\, p_A(z',t \, |\,  z).
\end{equation}
A similar relation can be derived to compute $p_A$ from~$p_H$; thus the solution for one boundary condition (BC) determines the solution for the other. Siegmund duality, first noted for Brownian motion~\cite{levyProcessusStochastiquesMouvement1992,lindleyTheoryQueuesSingle1952} and later formalized in~\cite{siegmundEquivalenceAbsorbingReflecting1976}, constitutes a particular instance of a Markov duality~\cite{cox1984duality, jansen2014notion, monthus2025markov}. At its core lies a time-reversal symmetry relating the forward Fokker-Planck operator  of a process to the backward operator of its dual~\cite{gueneauSiegmundDualityPhysicists2024}. Constructing the dual of a stochastic process is, however, generally a nontrivial task. 
Only recently have explicit constructions been obtained for continuous-time stochastic processes driven by an equilibrium Markov process, such as active particles, and discrete-time random walks with stationary time-reversal-symmetric increments, which may be correlated and non-Markovian~\cite{gueneauSiegmundDualityPhysicists2024,gueneauRelatingAbsorbingHard2024,gueneau:tel-05238622,touzo2025exact}

In this Letter, we derive a systematic expansion for the absorbing-wall propagator in the low-activity regime and analyze the high-activity regime in the long-time limit, providing access to the first-passage statistics. We then use these analytical results to determine the hard-wall propagator via Eq.~\eqref{eq:propagator_relation}, which we also prove for ABPs. 
Our analytical results show that, depending on the initial position and orientation, activity can either enhance or hinder time efficiency. Moreover, the dynamics between hard-walls relaxes to a wall-accumulated stationary state, akin to that found in many active systems~\cite{rothschildNonrandomDistributionBull1963,winetObservationsResponseHuman1984,berkeHydrodynamicAttractionSwimming2008,Denissenko:2012, elgetiWallAccumulationSelfpropelled2013,caprini_active_2018, caprini_active_2019, caprini_transport_2019}.

\par~\textit{Model.--} We consider the motion of a two-dimensional ABP moving at a constant speed $v$ along its instantaneous orientation 
$\mathbf{e}(t) = (\sin \vartheta(t), \cos \vartheta(t))$, which is subject to rotational diffusion with diffusivity $D_{\mathrm{rot}}$. The particle position also experiences translational diffusion with diffusivity~$D$. The position $\mathbf{r}(t)=(x(t),z(t))$ and orientation $\vartheta(t)$  are governed by the Langevin equations:
\begin{align}
\dot{\mathbf{r}} &= v\, \mathbf{e} + \sqrt{2D}\,\boldsymbol{\eta} \quad {\rm and} \quad 
\dot{\vartheta} = \sqrt{2D_{\mathrm{rot}}}\,\xi\, , \label{eq:langevin_theta}
\end{align}
where $\boldsymbol{\eta}(t)$ and $\xi(t)$ are independent, delta-correlated Gaussian white noises with zero mean. The ABP is confined between two walls located at $z=0$ and $z=L>0$ with two different BCs: \textit{absorbing (A)} and \textit{hard (H) walls}. Its dynamics are characterized by three timescales: the diffusive time $\tau=L^{2}/D$, the ballistic time $\tau_a = L/v$, and the rotational diffusion time $\tau_{\mathrm{rot}}=1/D_{\mathrm{rot}}$. This entails two nondimensional numbers: the P{\'e}clet number $\mathrm{Pe}=\tau/\tau_a=vL/D$, measuring active motion relative to diffusion, and $\gamma=\tau/\tau_{\mathrm{rot}}=D_\mathrm{rot} L^2/D$, reflecting the importance of rotational {\it vs.} translational diffusion. 

Since we are only interested in the dynamics perpendicular to the walls, we focus on the $z-$coordinate. We denote the associated propagators, averaged over the equilibrium initial orientation, i.e. $p_{\rm eq}(\vartheta_0) = 1/(2\pi)$, and integrated over the final orientation, by 
\begin{eqnarray}
    \!\!\!\!\!\!\!\! p_{A,H}(z,t \, |\,  z_{0}) = \int {\rm d}\vartheta \int \frac{{\rm d}\vartheta_0}{2\pi}\, p_{A,H}(z,\vartheta,t \, |\, z_0, \vartheta_0)\, ,
\end{eqnarray}
where $p_{A,H}(z,\vartheta,t \, |\, z_0, \vartheta_0) {\rm d}z\, {\rm d}\vartheta$ is the probability that the ABP is found in the interval $[z, z+dz]$ with orientation in $[\vartheta,\vartheta+{\rm d}\vartheta]$ at time $t$, given the initial conditions $z(t=0)=z_0$ and $\vartheta(t=0)= \vartheta_0$. 
We first demonstrate Eq.~\eqref{eq:propagator_relation} by adapting the main steps of Ref.~\cite{gueneauSiegmundDualityPhysicists2024}. This, in turn, enables us to derive analytical results for the time-dependent distribution between hard walls from the absorbing-wall propagator of ABPs.

\paragraph{Absorbing (sticking) walls.--} We first consider an ABP with sticking absorption at $z=0$ and $z=L$:
upon first hitting either wall, the particle remains (`sticks') there for all later times. Accordingly, the propagator may contain
Dirac masses at $z=0$ and $z=L$ weighted by the time-integrated incoming boundary flux up to time $t$. It satisfies the backward
Fokker--Planck equation (FPE)
\begin{eqnarray}
&\partial_t p_A(z,t\, |\, z_0,\vartheta_0)
=
\mathcal L^\dagger_{z_0,\vartheta_0}\,p_A(z,t\, |\, z_0,\vartheta_0)\, ,\label{eq:BWFP}
\end{eqnarray}
where the backward operator $\mathcal L^\dagger_{z_0,\vartheta_0}:= v\cos\vartheta_0\,\partial_{z_0}
+ D\, \partial_{z_0}^2 + D_{\rm rot}\,\partial_{\vartheta_0}^2 $ acts on the \emph{initial} variables $(z_0,\vartheta_0)$, and we have averaged over the final orientation. We define the probability to find the ABP in $[\ell_A>0,L]$ at time $t$ with initial conditions~$(z_0,\vartheta_0)$:
\begin{subequations}
\begin{eqnarray}\label{eq:QA_def}
Q_A(\ell_A,t\, |\,z_0,\vartheta_0)
:&=&
\mathbb P_A\!\left(z(t)\ge \ell_A \, |\,z_0,\vartheta_0\right),\\
&=&
\int_{\ell_A}^{L}\!{\rm d}z\;p_A(z,t\, |\, z_0,\vartheta_0)\, .
\end{eqnarray}
\end{subequations}
Since $\mathcal L^\dagger$ acts only on the initial variables,
integrating Eq.~\eqref{eq:BWFP} over $z\in[\ell_A,L]$ yields
\begin{equation}\label{eq:QA_PDE}
\partial_t Q_A(\ell_A,t\, |\, z_0,\vartheta_0)
=
\mathcal L^\dagger_{z_0,\vartheta_0}\,Q_A(\ell_A,t\, |\, z_0,\vartheta_0)\, ,
\end{equation}
subject to BCs $Q_A(\ell_A,t\, |\,  0,\vartheta_0)=0$, since a particle starting at $z=0$ is absorbed immediately, and $Q_A(\ell_A,t\, |\,  L,\vartheta_0)=1$, as a particle starting at $z=L$ is absorbed and $z=L\ge \ell_A$. Moreover, from the initial condition   $p_A(z,0\ |z_0, \vartheta_0)=\delta(z-z_0)$, we obtain $Q_A(\ell_A,0\, |\,  z_0,\vartheta_0)= \Theta(z_0 -\ell_A)$ where $\Theta(x)$ is the Heaviside function.

\begin{figure*}[tp]
\includegraphics[width=1.0\textwidth]{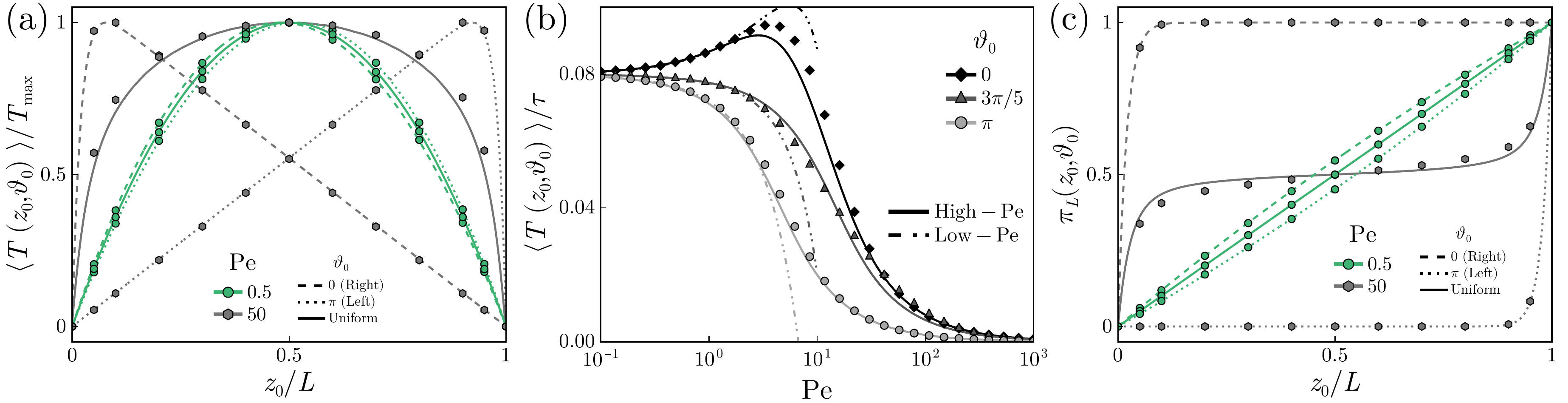}
\caption{\label{fig:2_mfpt_splitting_2} First-passage properties. (a)~Mean-first-passage time $\langle T \rangle$ normalized by its maximum $T_{\mathrm{max}}$, as a function of the initial position $z_{0}$. (b)~Mean-first-passage time $\langle T \rangle$ as a function of the P{\'e}clet number and for $z_{0}/L=0.2$. (c)~Splitting probability $\pi_{L}$ at the right wall as a function of $z_0$. In (a--c) we set $\gamma=3$. 
In (b), dash-dotted and solid lines denote the low-$\mathrm{Pe}$ expansion up to $O(\mathrm{Pe}^2)$ and the high-activity result [Eq.~\eqref{eq:mfpt_high_pe}], respectively. Symbols correspond to simulation results.
}
\end{figure*}

\paragraph{Hard (reflective) walls.--}
We next consider an ABP between two hard walls at $z=0$ and $z=L$ and denote by $ p_H(z,\vartheta,t\, |\, \tilde z_0)$ the probability density averaged over the initial angle. It satisfies the forward FPE
\begin{eqnarray}
&\partial_t  p_H(z,\vartheta,t\, |\, \tilde z_0)
=
\mathcal L_{z,\vartheta}\, p_H(z,\vartheta,t\, |\, \tilde z_0)\, , \label{eq:HWFP}
\end{eqnarray}
where the forward operator $\mathcal L_{z,\vartheta}:= -v\cos\vartheta\,\partial_{z}
+ D\, \partial_{z}^2 + D_{\rm rot}\,\partial_{\vartheta}^2$ acts on the \emph{final} variables~$(z,\vartheta)$. The system is subject to no-flux BCs:
$v\cos\vartheta\, p_H-D\, \partial_z  p_H=0$ at $z=0$ and $L$.
We now define the conditional probability to find the ABP in $[0, \ell_H]$ at time~$t$ given the fixed final orientation $\vartheta(t)=\vartheta$ as
\begin{subequations}
\begin{eqnarray}\label{condnitionalStep}
Q_H(\ell_H,t\, |\, \vartheta,\tilde z_0)
:&=&\mathbb P_H(z(t)\le \ell_H \,\big|\;  \vartheta,\tilde z_0), \quad \\
&=& \frac{ \int_{0}^{\ell_H}\!{\rm d}z\;  p_H(z,\vartheta,t\,\big|\,\tilde z_0)}{p_{\mathrm{eq}}(\vartheta)}\, .\label{condnitionalStepBayes}
\end{eqnarray}
\end{subequations}
Here,  Eq.~(\ref{condnitionalStepBayes}) follows from Bayes’ rule: $p_H(z,t \, | \, \vartheta,\tilde z_0)
= p_H(z,\vartheta,t  \, | \,  \tilde z_0)/p(\vartheta,t  \, | \,  \tilde z_0)$. As $\vartheta(t)$ evolves independently of $z(t)$ and is initialized at stationarity, $\vartheta(0)\sim p_{\mathrm{eq}}(\vartheta)$, its marginal remains $p_{\mathrm{eq}}(\vartheta)$ for all $t$ (hence $p(\vartheta,t\, | \, \tilde z_0)=p_{\mathrm{eq}}(\vartheta)$). Further, since $p_{\mathrm{eq}}(\vartheta)=1/(2\pi)$ is $\vartheta$-independent,  integrating Eq.~\eqref{eq:HWFP} over $z\in[0,\ell_H]$ yields
\begin{equation}\label{eq:QH_PDE}
\partial_t Q_H(\ell_H,t\, |\, \vartheta,\tilde z_0)
=
\mathcal L_{\ell_H,\vartheta}\,Q_H(\ell_H,t\, |\, \vartheta,\tilde z_0)\, .
\end{equation}
More generally, if $p_{\mathrm{eq}}(\vartheta)$ is not constant, Eq.~(\ref{eq:QH_PDE}) remains valid provided the $\vartheta$-dynamics is reversible, independent of $z(t)$, and initialized at stationarity~\cite{gueneauSiegmundDualityPhysicists2024}.
Translational Brownian noise prevents singular boundary masses. Therefore, the probability to lie in $[0,\ell_H]$ vanishes at $\ell_H=0$, yielding the BC $Q_H(0,t\, | \,\vartheta, \tilde z_0)=0$. 
Further, hard-wall confinement enforces $z(t)\in[0,L]$ for all~$t$, hence $Q_H(L,t\,|\,\vartheta,\tilde z_0)=1$, and the localized start $ p_H(z,\vartheta,0\,|\,\tilde z_0)=\delta(z-\tilde z_0)$ translates to $Q_H(\ell_H,0\, |\, \vartheta,\tilde z_0)= \Theta(\ell_H - \tilde{z}_0)$.

\paragraph{Siegmund duality.--}
Under the angle shift $\vartheta=\vartheta_0+\pi$, which encodes the reversal of the active drift $-v\cos(\vartheta_0+\pi)=v\cos\vartheta_0$, the operators 
match $\mathcal L_{z_0,\vartheta_0+\pi}=\mathcal L^\dagger_{z_0,\vartheta_0}$, i.e., the hard-wall process at time~$t$ is conditioned on an orientation opposite to the absorbing process’s initial one. Identifying $\ell_A=\tilde z_0$ in Eq.~\eqref{eq:QA_def} and $\ell_H= z_0$ in Eq.~(\ref{condnitionalStep}), their initial conditions also coincide. Given these constraints, $Q_H(z_0,t\, |\, \vartheta(t)=\vartheta_0+\pi,\tilde z_0)$ and
$Q_A(\tilde z_0,t\, |\, z_0,\vartheta_0)$ satisfy the same differential equation with the same boundary and initial conditions, and are therefore equal.
Averaging over $\vartheta_0\sim p_{\mathrm{eq}}$ and using the invariance of $p_{\mathrm{eq}}$ under
$\vartheta_0\mapsto \vartheta_0+\pi$ yields the Siegmund duality relation~\cite{siegmundEquivalenceAbsorbingReflecting1976, gueneauSiegmundDualityPhysicists2024}
\begin{equation}\label{eq:siegmund_avg}
\mathbb P_H(z(t)\le z_0 \,\big|\; \tilde z_0)
=
\mathbb P_A(z(t)\ge \tilde z_0 \,\big|\; z_0)\, .
\end{equation}
Hence, after averaging the initial orientations over $p_{\text{eq}}(\vartheta)$, the probability that the hard-wall ABP started at $\tilde z_0$ lies to the left of $z_0$ at time $t$ equals the probability that the absorbing ABP started at $z_0$ lies to the right of $\tilde z_0$ at the same time. Differentiating Eq.~\eqref{eq:siegmund_avg} with respect to $z_0$, and renaming variables, straightforwardly gives the announced result in Eq.~\eqref{eq:propagator_relation}. 

The duality relation [Eq.~\eqref{eq:siegmund_avg}] also allows deriving the survival probability from the hard-wall propagator via (see End Matter)
\begin{eqnarray}\label{survivalSiegmund}
        S(z_0, t) = \int_0^{z_0}dz \left[p_H(z,t|0) - p_H(z,t|L)\right]\, , 
\end{eqnarray}
and relates the mean-first-passage time (MFPT) from $z_0$ to either wall to the local time spent by ABPs in~$[0,z_0]$ between hard walls: 
\begin{eqnarray}\label{MFPTSiegmund}
    \langle T \rangle = \int_0^{z_0}dz \int_0^{\infty}dt\left[p_H(z,t|0) - p_H(z,t|L)\right]\, .
\end{eqnarray}
Interestingly, Eqs.~(\ref{survivalSiegmund}) and ~(\ref{MFPTSiegmund}) are reminiscent of the relations derived in Ref.~\cite{levernier_survival_2019}, where first-passage observables are connected to time integrals of propagator differences in complementary geometries.

\begin{figure*}[tp]
\includegraphics[width=0.95\textwidth]{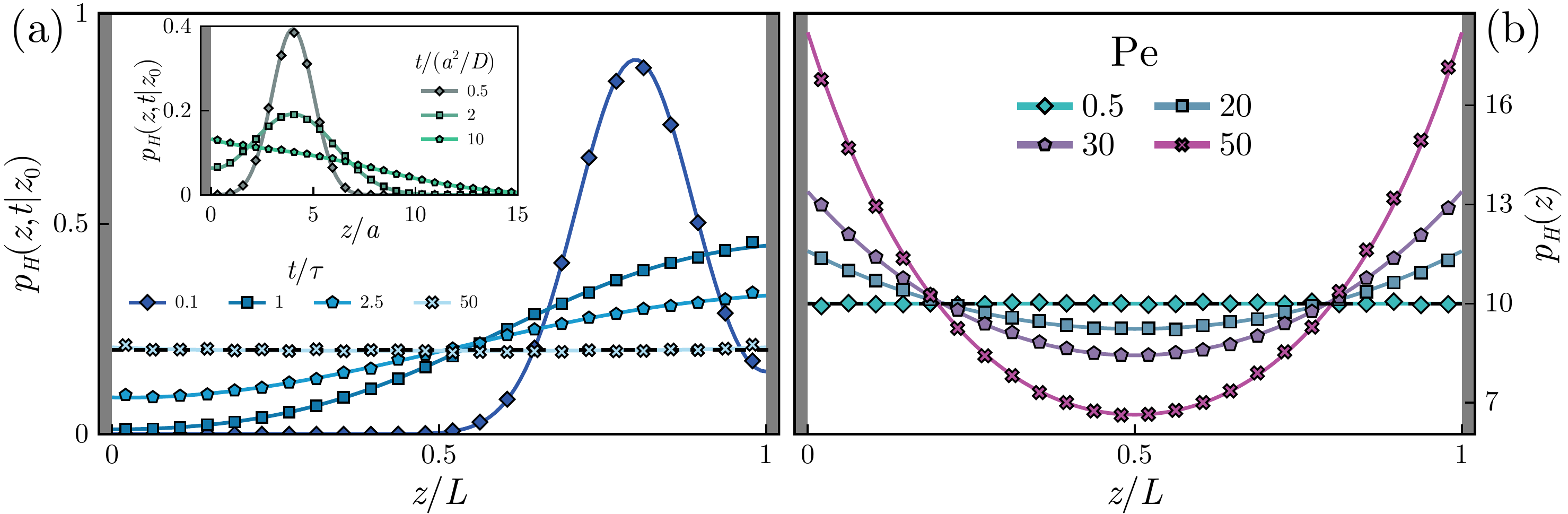}
\caption{\label{fig:3_propagator} Probability densities of an ABP between hard walls. (a)~Probability density $p_{H}(z,t|z_0)$ as a function of the position $z$ for different times $t$. Here, $z_{0}/L =0.8$ and $\mathrm{Pe}=0.5$. Lines correspond to the low-$\mathrm{Pe}$ (up to $O(\mathrm{Pe}^{2})$) expansion and symbols are simulation results. (\textit{Inset}) Probability density $p_{H}(z,t|z_0)$ for the case of a single wall at $z=0$.  We set $z_{0}/a =4$ and $\mathrm{Pe}=0.5$, and rescale length scales with the particle's hydrodynamic radius $a = \sqrt{3D/(4D_{\mathrm{rot}})}$. (b)~Stationary distribution $p_{H}(z)$ for different P{\'e}clet numbers. Lines correspond to the high-$\rm Pe$ solution~\eqref{eq:pH_stationary} and symbols are simulation results. (a--b)~Vertical gray lines indicate~the~walls. }
\end{figure*}

\par~\textit{First-passage properties.--} 
We decompose the propagator of an ABP between absorbing walls $p_A(z,  t|z_0)$ as
\begin{equation}\label{PDFexit}
    p_A = {{P_A}} + E_0\ \delta(z) + E_L\ \delta(z-L)\, ,
\end{equation}
where $P_A(z,t|z_{0})$ denotes the probability density of an ABP between absorbing non-sticking walls, i.e., boundaries for which the particle disappears upon reaching a wall ($P_A(z=0,t) = P_A(z=L,t)=0$), while boundary masses are encoded in the exit probabilities:
\begin{equation}\label{fluxExit}
    E_{0,L}(z_0, t)
= \pm D\int_0^t {\rm d}t'\, 
\partial_z P_A(z,t'|z_0)\big|_{z=0,L}\, .
\end{equation}
In the low activity regime (${\rm Pe}\ll 1$), where translational diffusion dominates over active motion, 
we compute $P_A(z,t|z_{0})=\int \mathrm{d}\vartheta\int \mathrm{d}\vartheta_{0} {{P_A}}(z,\vartheta,t|z_{0},\vartheta_0)/(2\pi)$ by solving the forward FPE [Eq.~\eqref{eq:HWFP}] with initial condition~$P_A(z,\vartheta,t=0| z_0,\vartheta_0)=\delta(z-z_{0})\delta(\vartheta-\vartheta_{0})$.  The time-dependent behavior can be obtained by first moving to Laplace space ($\widehat{f}(s) = \int_0^\infty f(t)e^{-st} \mathrm{d}t $) and inserting the P{\'e}clet number expansion $P_A=\widehat{P}_A^{(0)}+ \mathrm{Pe}~\widehat{P}_A^{(1)} + \mathrm{Pe}^{2}\widehat{P}_A^{(2)} + O(\mathrm{Pe}^{3})$ into Eq.~\eqref{eq:HWFP}. Here, $\widehat{P}_A^{(0)}$ corresponds to the propagator of a diffusive particle (with rotational diffusion) between two absorbing walls and can be readily computed using the method of reflections (see End Matter). This procedure yields a recursive relation
\begin{align}
    &\widehat{P}_A^{(i+1)}(z,\vartheta,s |z_{0}, \vartheta_{0}) =  \int_{0}^{L} \mathrm{d}z'  \int_{0}^{2\pi} \mathrm{d}\vartheta' ~ \widehat{P}_A^{(0)}(z, \vartheta, s|z', \vartheta') \notag\\& \qquad \times \left[- (D/L)\cos(\vartheta')\partial_{z'} \widehat{P}^{(i)}_A\right]( z', \vartheta',s | z_{0},\vartheta_{0})\, . \label{eq:integral_perturbation}
\end{align}
We then use the survival probability $\widehat{S}(s|z_{0}, \vartheta_{0})= \int_{0}^{2\pi}\int_{0}^{L}\widehat{P}_A(z, \vartheta,s|z_{0}, \vartheta_{0}) \mathrm{d}z \mathrm{d} \vartheta$ 
to compute the MFPT via $\langle T\rangle= \widehat{S}(s=0 |z_{0}, \vartheta_{0})$ [Fig.~\ref{fig:2_mfpt_splitting_2}].

Analytical progress is also possible in the high-activity regime, where persistent motion  dominates over orientational diffusion $\tau_a \ll \tau_{\rm rot}$ (i.e., $\gamma \ll {\rm Pe}$). Therefore, we introduce $\epsilon=\gamma/{\rm Pe}$ and first look into the boundary layer of width $O(L\epsilon)$. Rescaling the equation for the MFPT, $\mathcal{L}^{\dagger}_{z_{0}, \vartheta_{0}}\langle T\rangle=-1$, with the stretched coordinate~$y=(z_{0}-L)/(L\epsilon)$ shows that (close to the walls) rotational diffusion can be neglected, given $\gamma \ll \mathrm{Pe}^{2/3}$ (see End Matter). Neglecting rotational diffusion reduces the high-activity limit to a standard one-dimensional biased-diffusion problem. Taking the BCs $\langle T  (0,\vartheta_{0})\rangle =\langle T(L,\vartheta_{0}) \rangle =0$, we obtain (for $\cos \vartheta_0 \neq 0$)~\cite{rednerGuideFirstpassageProcesses2001}:
\begin{align}
    \langle T \rangle \underset{\gamma/\mathrm{Pe} \to0}{=} \frac{\tau}{{\rm Pe}\cos\vartheta_0}\left[\frac{1-e^{-{\rm Pe}\cos(\vartheta_0) z_0/L}}{1-e^{-\mathrm{ Pe}\cos(\vartheta_0)}}-\frac{z_0}{L}\right]  \label{eq:mfpt_high_pe}
\end{align}
Our results [Fig.~\ref{fig:2_mfpt_splitting_2}(a)] show that randomly-oriented, active particles take the longest when starting at the channel's midpoint, with a symmetric MFPT w.r.t. the initial position $z_0$. 
This picture changes for particles initially oriented towards the right wall ($z/L=1$): At low activity ($\mathrm{Pe} =0.5$), the maximal MFPT shifts to $z_0/L\lesssim 0.5$ and becomes asymmetric, because active motion helps the particle to reach the right wall (towards which it is oriented) faster. For large P{\'e}clet numbers ($\mathrm{Pe} =50$), the peak of the MFPT shifts towards $z_{0}\gtrsim0$ and decays linearly thereafter, which reflects the typical time taken by a persistent particle to cover the remaining distance at constant speed: $\langle T\rangle \simeq (L-z_{0})/v$. The picture reverses when the particle initially travels towards the left wall.

We further investigate the MFPT as a function of $\mathrm{Pe}$, which decreases monotonically for particles starting at $z_0=0.2L$ and oriented towards the closer wall at $z=0$ ($\vartheta_0=3\pi/5, \pi$) [Fig.~\ref{fig:2_mfpt_splitting_2}(b)], showing that active motion speeds up their wall arrival. Interestingly, the MFPT becomes non-monotonic for particles starting close to a wall and departing towards the opposite one ($\vartheta_0=0$). While the agent reaches the closer boundary through diffusion at $\mathrm{Pe} \lesssim1$ with $\langle T\rangle\simeq z_{0} (L-z_{0})/(2D)$, it reaches the opposite boundary almost instantaneously for $\mathrm{Pe} \gtrsim 100$. 
At intermediate Pe the MFPT increases, which can be understood from the small-Pe expansion of Eq.~\eqref{eq:mfpt_high_pe}: $
\langle T \rangle / \tau = z_0(L-z_0)/(2L^2) + C(z_0)\cos\vartheta_0\,\mathrm{Pe} + O(\mathrm{Pe}^2)$, where $C(z_0=0.2L)>0$. Thus, the MFPT increases  for particles oriented towards the opposite wall $|\vartheta_0|<\pi/2$, leading to an overall non-monotonic behavior as the MFPT decreases as $\sim 1/{\rm Pe}$ at large Pe. Physically, this behavior results from the fact that active motion takes the particle away from the closer wall and thus increases the time to reach the latter via translational diffusion. We note that Eq.~(\ref{eq:mfpt_high_pe}) agrees  with our simulations at large $\mathrm{Pe} \gg \gamma^{3/2}\approx 5.2$ and low $\mathrm{Pe}$, while discrepancies appear at moderate Pe and at larger $\gamma$, where rotational diffusion becomes important. Our low-$\mathrm{Pe}$ expansion captures the simulations irrespective of $\gamma$ and notably until $\mathrm{Pe} \approx 10$ for selected cases, as it accounts for the rotational-translational coupling (see End Matter, Fig.~\ref{fig:mfpt_pe_gamma}~(a,b)).

Another important first-passage quantity is the splitting probability~\cite{rednerGuideFirstpassageProcesses2001, klingerSplittingProbabilitiesSymmetric2022, majumdarHittingProbabilityAnomalous2010, iyaniwura_splitting_2026}, i.e., the probability that the agent reaches the right wall before the left one. For small~{\rm Pe}, it can be computed as the long-time limit of the exit probability $\pi_L(z_0) = E_L(z_0,t\to\infty)$ [Eq.~\eqref{fluxExit}]. For high activity, where the dynamics reduce to biased diffusion, we employ the same approach as for~$\langle T\rangle$ (with BCs $\pi_{L}(0,\vartheta_{0})= 0$, $\pi_{L}(L,\vartheta_{0})=1$), leading to~\cite{rednerGuideFirstpassageProcesses2001}: 
\begin{align}\label{splittingletter}
    \pi_{L}(z_{0}, \vartheta_{0}) \underset{\gamma/\mathrm{Pe} \to 0}{=} \frac{1-e^{-\mathrm{Pe} \cos(\vartheta_{0})z_{0}/L}} {1-e^{- \mathrm{Pe}\cos(\vartheta_{0})} }.
\end{align}
See also Ref.~\cite{iyaniwura_splitting_2026} for a related high-activity boundary-layer analysis of the splitting probability.
For randomly-oriented agents, the splitting probability at the right wall~$\pi_L$ increases almost linearly with $z_{0}$ at low activity, where translational diffusion dominates [Fig.~\ref{fig:2_mfpt_splitting_2}(c)]. Our results further show that particles initially oriented towards the right wall have a higher probability of reaching it than those oriented away from it. The difference in probabilities becomes smaller at the boundaries, where translational diffusion outperforms active motion. At high activity, however, $\pi_L$ approaches a plateau at $\approx0.5$ for most $z_{0}$, reflecting the ballistic motion of the particle towards either wall with equal probability. For a fixed initial orientation, the behavior drastically changes: $\pi_L$ jumps to $1$ when $\vartheta_0$ points towards the wall at $z=L$, and to $0$ when it points towards the opposite one.

\par~\textit{Confined particles.--}
Leveraging Siegmund's duality [Eq.~\eqref{eq:propagator_relation}] and Eq.~(\ref{eq:integral_perturbation}), we compute the time-dependent propagator in the presence of hard-walls~$p_{H}(z,t|z_0)$. 
Setting $\tilde z_0 = L$ in Eq.~\eqref{eq:siegmund_avg} and taking $t \to \infty$ shows that the stationary distribution $p_{H}(z) = p_{H}(z,t \to \infty|z_0)$ is a simple derivative of the splitting probability~\cite{gueneauRelatingAbsorbingHard2024,gueneauSiegmundDualityPhysicists2024} (see End Matter). In the high-activity regime, it can thus be computed from Eq.~(\ref{splittingletter}), without solving the stationary FPE directly. It reads
\begin{equation} \label{eq:pH_stationary}
    \!\!\!\!p_{H}(z) \!=\! \frac{\partial\pi_{L}}{\partial z} \!=\! \frac{\rm Pe}{2\pi L}\int_0^{2\pi} \!\!\cos(\vartheta_{0}) \frac{e^{-{\rm Pe} \cos(\vartheta_{0})z/L} } {1-e^{- \mathrm{Pe} \cos(\vartheta_{0})}}\mathrm{d}\vartheta_{0}\, ,
\end{equation}
thereby recovering the result of Ref.~\cite{elgetiWallAccumulationSelfpropelled2013}. Figure~\ref{fig:3_propagator}~(a) shows how, starting from a localized initial condition, the probability density spreads over the channel width and eventually reaches a stationary distribution. The latter evolves from an almost uniform profile at low activity ($\mathrm{Pe}=0.5$) to an increasingly pronounced U-shaped profile as activity increases [Fig.~\ref{fig:3_propagator}(b)], indicating enhanced wall accumulation. In our model, this accumulation results solely from the interactions of active agents with hard walls, as active particles remain near the boundary until rotational diffusion allows them to reorient and escape. This finding is in qualitative agreement with  experimental~\cite{rothschildNonrandomDistributionBull1963,winetObservationsResponseHuman1984,berkeHydrodynamicAttractionSwimming2008} and computational~\cite{elgetiWallAccumulationSelfpropelled2013} observations, yet the  importance of steric {\it vs.} hydrodynamic interactions~\cite{berkeHydrodynamicAttractionSwimming2008}, yielding similar wall-accumulated distributions, may depend on the specific systems. 

The time-dependent propagator with hard walls (in the low-Pe regime) provides direct access to the moments of the position $\langle z^{n}(t) \rangle = \int_{0}^{L} p_{H}(z,t|z_0) z^{n} \mathrm{d}z$. As suggested by the stationary distribution, the mean position converges to the channel's midpoint, $\langle z(t) \rangle \underset{t\to\infty}{=}  L/2 $, while the fluctuations around it are perturbed $\langle (z(t)-\langle z(t)\rangle)^2\rangle \underset{t\to\infty}{=} (L^2/12) + A\mathrm{Pe}^{2}+ O(\mathrm{Pe}^{4})$ with~$A>0$, showing that activity enhances the uniform Brownian result. 

Finally, our model recovers the one-wall problem in the limit of infinite channel width, $L\to\infty$ [Fig.~\ref{fig:3_propagator}(a)({\it inset})]. In this case, the MFPT diverges~\cite{baoucheFirstpassagetimeStatisticsActive2025} and no stationary distribution exists: without the right boundary, the particle can escape to infinity, leading to a probability density that broadens continuously in space over time.

\par~\textit{Conclusions.--} 
We have analytically characterized active Brownian dynamics between two walls. A systematic low-Péclet expansion derived for the absorbing problem can be transferred through Siegmund duality to the time-dependent probability density between hard walls. We thereby quantified the evolution towards a stationary wall-accumulated state, thus providing a theoretical footing for experimental and computational observations of active agents. Our work adds to our understanding of active Brownian transport in channels, by unraveling how the interplay of active motion and diffusion dictates first-passage statistics of active agents between two walls.

The formalism employed here offers a systematic approach for accessing both time-related efficiencies and spatial distribution properties, in a class of problems that are otherwise difficult to address. It admits several extensions: these include alternative agent dynamics, such as run-and-tumble motion~\cite{Cates:2013, Kurzthaler:2024, gueneauRelatingAbsorbingHard2024,gueneau_exact_2026, frydel_run-and-tumble_2024}, stochastic resetting processes~\cite{evans_diffusion_2011, Evans:2020, Baouche:2025, gueneauSiegmundDualityPhysicists2024, kumar_active_2020}, non-Markovian processes~\cite{levernier_survival_2019, arnoulx_de_pirey_exact_2024, gueneauSiegmundDualityPhysicists2024}, higher-dimensional settings that could introduce non-trivial modifications~\cite{Straube_Höfling_2024}, more complex geometries, e.g., structured walls~\cite{zhaoActiveParticlesTube2025, Malgaretti:2017,Granek:2024, iyaniwura_splitting_2026} or concentric confinement, and partially-absorbing boundaries~\cite{schumm_search_2021,bressloff_trapping_2023,bressloff_run-and-tumble_2025}. The latter problems may prove more tractable in one setting -- absorbing or reflecting -- whose solution can be systematically transferred to its dual counterpart. Beyond confining boundaries, this framework may be applied to study the impact of potential barriers on escape problems~\cite{Woillez_Zhao_Kafri_Lecomte_Tailleur_2019, gueneauRunandtumbleParticleOnedimensional2025}. Overall, it may thus enable progress in quantifying biological and soft-matter systems, where confinement strongly regulates transport and reaction efficiency, while systematic theoretical approaches remain comparatively scarce.

\vspace*{0.3cm}
\noindent {\it Acknowledgments.--} M.G. thanks L.~Touzo for insightful discussions on Siegmund duality and for past collaborations on this topic. We further thank Thomas Franosch for helpful discussions.
\bibliography{bibliography}

\appendix
\section{End Matter}
\paragraph{Perturbation expansion.--}
The probability density of an ABP between absorbing walls $P_A(z,\vartheta, t|z_0, \vartheta_0)$ satisfies the FPE: 
 $\partial_{t}P_{A} = \mathcal{L}_{z, \vartheta}P_{A} 
    =: \left[ \mathcal{H}_{0}+ \mathrm{Pe}\mathcal{V} \right]P_{A}\,$,
where $\mathcal{H}_0=D\partial_z^2+D_{\mathrm{rot}}\partial_\vartheta^2$ and
$\mathcal{V}=- (D/L)\cos\vartheta\,\partial_z$. Transforming to Laplace space  and inserting  $\widehat{P}_{A}\!=\!\widehat{P}_{A}^{(0)}\!+\! \mathrm{Pe}~\widehat{P}_{A}^{(1)} \!+\! \mathrm{Pe}^{2}\widehat{P}_{A}^{(2)} \!+\! \cdots$ yields the  coupled equations:
\begin{subequations}
    \begin{align}
        (s-\mathcal{H}_{0}) \widehat{P}_{A}^{(0)} &=\delta(z-z_{0})\delta(\vartheta-\vartheta_{0}), \label{eq:iterative0} \\
        (s-\mathcal{H}_{0}) \widehat{P}_{A}^{(i+1)} &= \mathcal{V} \widehat{P}_{A}^{(i)}, \qquad i\ge0\label{eq:iterative1}\, ,
    \end{align}
\end{subequations}
where the zeroth order $P_{A}^{(0)}$ corresponds to the propagator of a passive particle with rotational diffusion between two absorbing walls. It is possible to perform an eigenfunction expansion of the unperturbed FPE and to obtain the Green’s function of the unbounded problem:
\begin{equation}
    G^{u}(s,z, \vartheta, z', \vartheta') =  \frac{1}{4\pi}\sum_{\ell=-\infty}^{\infty} \frac{1}{ Dp_{\ell}}e^{-p_{\ell}|z-z'|} e^{i \ell (\vartheta' -\vartheta)}\, ,
\end{equation}
where $p_{\ell}^{2} =  \left(s+ D_{\mathrm{rot}}\ell^{2} \right)/D$. 
The absorbing boundaries at $z=0$ and $z=L$ are then enforced by the method of reflections, which gives
\begin{widetext}
\begin{equation}
\widehat{P}^{(0)}_A(z,\vartheta,s\mid z',\vartheta')=
\frac{1}{2\pi}
\sum_{\ell=-\infty}^{\infty}
e^{i\ell(\vartheta' - \vartheta)}
\frac{1}{Dp_\ell \sinh \left( p_{\ell} L\right)}
\begin{cases}
\sinh\!\left(p_\ell z\right)
\,\sinh\!\left(p_\ell\left(L-z'\right)\right),
& z < z', \\[1.2ex]
\sinh\!\left(p_\ell z'\right)
\,\sinh\!\left( p_\ell\left(L-z\right)\right),
& z > z'.
\end{cases}
\label{P0piecewise}
\end{equation}
 \end{widetext}
Equations~(\ref{eq:iterative0}) and~(\ref{eq:iterative1}) lead to the iterative integral relation in Eq.~(\ref{eq:integral_perturbation}), which we use to compute each order of the perturbative expansion. Note that this method is adapted from Ref.~\cite{baoucheFirstpassagetimeStatisticsActive2025}, where the case of a single absorbing wall was studied.
\begin{figure*}[ht]
    \centering
    \includegraphics[width=\linewidth]{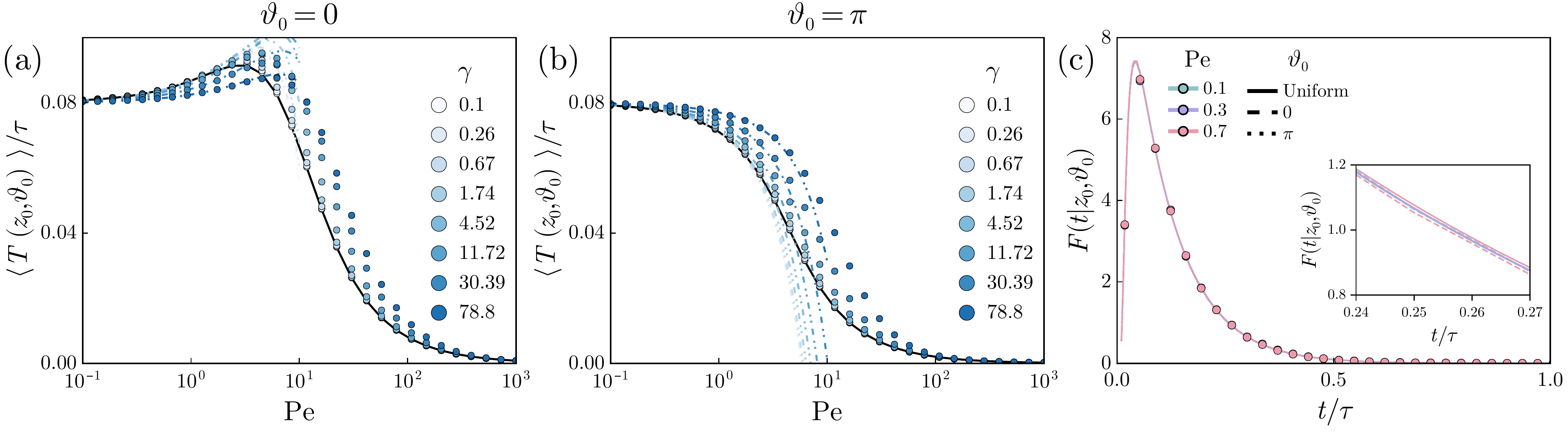}
    \caption{(a-b) Mean-first-passage time $\langle T \rangle$ as a function of the P{\'e}clet number for different $\gamma$ and for $z_{0}/L=0.2$. Dash-dotted and solid lines denote the low-$\mathrm{Pe}$ expansion up to $O(\mathrm{Pe}^2)$ and the high-activity result, respectively. (c) FPT~probability density $F(t|z_0,\vartheta_0)$  as a function of time for different P{\'e}clet numbers and initial orientations, with initial position $z_0/L =0.5$ and $\gamma=3$. Inset shows a zoom of $F(t|z_0,\vartheta_0)$ at short times.}
    \label{fig:mfpt_pe_gamma}
\end{figure*}
The explicit corrections to the propagator are lengthy and not reported here (see the Mathematica Script~\footnote{The file is available on \href{https://doi.org/10.5281/zenodo.18982482}{Zenodo~(DOI: 10.5281/zenodo.18982482)}}). Instead, we only quote the first corrections to the main observables. 
\begin{widetext}
For the mean first-passage time, we obtain 
\begin{equation}
    \frac{\langle T(z_0,\vartheta_0) \rangle}{\tau} = \frac{z_{0}\left( 1 - z_{0}/L\right)}{2L} + \cos\vartheta_0\frac{
 \sinh\!\left(\sqrt{\gamma}\left(z_0/L-\frac{1}{2}\right)\right)
-2\left(z_0/L-\frac{1}{2}\right)\sinh\!\left(\frac{1}{2}\sqrt{\gamma}\right)
}{
2\gamma \sinh\!\left(\frac{1}{2}\sqrt{\gamma}\right)
}\mathrm{Pe}
+ O(\mathrm{Pe}^2)\, .
\end{equation}
The splitting probability with fixed initial orientation reads
\begin{equation}
\begin{aligned}
\pi_L(z_0,\vartheta_0)
= \frac{z_0}{L}  + \cos\vartheta_0 \frac{
\cosh\!\left(\tfrac{1}{2}\sqrt{\gamma}\,\right)
-
\cosh\!\left(\sqrt{\gamma}\left(z_0/L-\tfrac{1}{2}\right)\right)
}{\gamma\,\cosh\!\left(\tfrac{1}{2}\sqrt{\gamma}\right)
} \mathrm{Pe}+ O(\mathrm{Pe}^2)\, .
\end{aligned}
\end{equation}
Finally, averaging over the initial orientation, we find (in agreement with Ref.~\cite{iyaniwura_splitting_2026})
\begin{equation}
\begin{aligned}
\pi_L(z_0) = \frac{1}{2\pi} \int_0^{2\pi}d\vartheta_0\, \pi_L(z_0,\vartheta_0) 
=  \frac{z_0}{L} + \frac{
\sinh\!\left(\sqrt{\gamma}\left(z_0/L-\frac{1}{2}\right)\right)
-2\left(z_0/L-\frac{1}{2}\right)\sinh\!\left(\tfrac{1}{2}\sqrt{\gamma}\right)
}{
2\gamma^{3/2} \cosh\!\left(\tfrac{1}{2}\sqrt{\gamma}\right)
} \mathrm{Pe}^2+ O(\mathrm{Pe}^4)\, .
\end{aligned}
\end{equation}
\end{widetext}

\paragraph{Consequences of Siegmund duality.--} From the Siegmund identity~[Eq.~\eqref{eq:siegmund_avg}], one can relate first-passage observables of the dynamics with absorbing walls to spatial properties of the dual dynamics with hard walls. Although illustrated here for ABPs, these relations hold generally for any pair of Siegmund-dual processes.

First, the survival probability of the ABP between two absorbing walls at $z=0$ and $z=L$ is defined by
\begin{subequations}
\begin{align}\label{}
\!\!\!\!\!S(z_0, &t) = \mathbb P_A(0< z(t)<  L \,\big|\; z_0)\, , \\
&= 1 - \mathbb  P_A(z(t)=  0 \,\big|\; z_0) - \mathbb  P_A(z(t)= L \,\big|\; z_0)\, ,  \\
&=\mathbb  P_A(z(t) \geq 0^+ \,\big|\; z_0) - \mathbb  P_A(z(t)\geq L \,\big|\; z_0) \, ,
\end{align}
\end{subequations}
where, in the last equality, we used the presence of an absorbing wall at $L$ and that $1 - \mathbb  P_A(z(t)=  0 \,\big|\; z_0) =\mathbb  P_A(z(t) \geq 0^+ \,\big|\; z_0) $. Using the Siegmund duality [Eq.~\eqref{eq:siegmund_avg}], it follows that
\begin{eqnarray}
        \!\!\!\!\!\!S(z_0, t) &=& \mathbb  P_H(z(t) \leq z_0 \,\big|\; 0^+) - \mathbb  P_H(z(t)\leq z_0\,\big|\; L)\, ,
\end{eqnarray}
yielding Eq.~(\ref{survivalSiegmund}) of the main text. In particular, in the presence of one wall (i.e., when $L \to \infty$), we find
\begin{equation}
    S(z_0, t)= \int_0^{z_0}{\rm d}z \, p_H(z,t|0)\, .
\end{equation}
Hence, the survival probability of a particle initially at~$z_0$ with an absorbing boundary at the origin can be interpreted as the probability that a dual particle, started at the origin and evolving in the presence of a hard wall at the origin, is located in the interval $[0,z_0]$ at time $t$.

Second, the MFPT follows directly from the survival probability, $\langle T \rangle = \int_0^{\infty}{\rm d}t\, S(z_0, t)$.
Using Eq.~(\ref{survivalSiegmund}) leads straightforwardly to Eq.~(\ref{MFPTSiegmund}). The MFPT to either absorbing wall for a particle starting at $z_0$ can thus be interpreted as the difference between the expected local times accumulated in the interval $[0,z_0]$ by the dual hard-wall dynamics when initialized at $0$ and~$L$.

Finally, setting $\tilde z_0 = L$ in~Eq.~\eqref{eq:siegmund_avg} relates the exit probability at wall $L$, defined in Eq.~(\ref{fluxExit}) for a particle starting at $z_0$, to the probability that the dual hard-wall dynamics lies in $[0,z_0]$~\cite{gueneauSiegmundDualityPhysicists2024},
\begin{equation}
    E_L(z_0,t) = \int_0^{z_0} {\rm d}z\, p_H(z,t\, | \, L)\, .
\end{equation}
In the long-time limit $t\to\infty$, the initial condition is forgotten. Taking a derivative with respect to $z_0$ then yields a direct relation between the splitting probability and the stationary distribution of the hard-wall dynamics~\cite{gueneauSiegmundDualityPhysicists2024},
\begin{equation}\label{splittingSS}
    p_H(z) = \frac{\partial \pi_L(z)}{\partial z}\, ,
\end{equation}
where $\pi_L(z)=\lim_{t\to\infty} E_L(z,t)$.

\paragraph{Boundary layer analysis for the MFPT.--}
The MFPT obeys the backward FPE: $\mathcal L^\dagger_{z_0,\vartheta_0}\langle T\rangle=-1 $. Away from the boundaries, the frozen-orientation approximation is justified when the rotational diffusion time is large compared to the persistent
crossing time, $L/v\ll D_{\rm rot}^{-1}$, or equivalently
$\gamma\ll{\rm Pe}$.
The condition is more restrictive inside the boundary layer close to the
walls. Near $z=L$, we introduce the stretched coordinate
$y=(z_0-L)/(L\epsilon)$ with $\epsilon=\gamma/{\rm Pe}$, which resolves a
layer of thickness $O(L\epsilon)$. The backward FPE then becomes
\begin{align}
  \left(
  \frac{D}{L^2\epsilon^2}\partial_y^2
  +\frac{v}{L\epsilon}\cos\vartheta_0\,\partial_y
  +D_{\rm rot}\,\partial_{\vartheta_0}^2
  \right)\langle T\rangle
  =
  -1 .
  \label{MFPT_BL_dim}
\end{align}
Multiplying by $\epsilon^2L^2/D$, using the definitions of ${\rm Pe}$ and $\gamma$, and rewriting ${\rm Pe}\,\epsilon=\gamma$, we arrive at
\begin{align}
  \left(
  \partial_y^2
  +\gamma\cos\vartheta_0\,\partial_y
  +\gamma\epsilon^2\,\partial_{\vartheta_0}^2
  \right)\frac{\langle T\rangle}{\tau}
  =
  -\epsilon^2\, .
\end{align}
Thus rotational diffusion is negligible in the boundary-layer equation provided $\gamma\epsilon^2\ll 1$, or  $\gamma\ll{\rm Pe}^{2/3}$.

\paragraph{Impact of $\gamma$ on the MFPT.--} The MFPT depends on rotational diffusion via $\gamma$. For $\vartheta_0=0$, an increase in $\gamma$ shifts the location of the peak towards larger Pe [Fig.~\ref{fig:mfpt_pe_gamma}(a)]. This can be understood by considering the particle's persistence length: $\ell_p=v/D_{\rm rot}=L{\rm Pe}/\gamma$. The largest MFPT is expected when the persistence length is comparable to (or shorter than) the distance to the wall at $z=L$: $\ell_p \sim L-z_0$, giving a critical P{\'e}clet number ${\rm Pe}^\star \sim \gamma(L-z_0)/L$ that increases with~$\gamma$. We further find that at small Pe, the MFPT decreases for large $\gamma$, indicating that rotational diffusion reorients particles towards the closer wall and thus speeds up their arrival. For large Pe, the MFPT increases with $\gamma$, as rotational diffusion perturbs the persistent motion towards the other wall (at $z=L$). This is accompanied by  non-monotonic behavior of the maximal MFPT w.r.t.~$\gamma$: Weak rotational diffusion (small $\gamma$) enhances the maximal MFPT compared to the high-activity limit, while the MFPT decreases at large~$\gamma$ and the peak smears out.

This effect is only prominent for particles oriented towards the wall, which is further away ($\vartheta_0=0$), while for particles oriented towards the closer wall ($\vartheta_0=\pi$) rotational noise always increases the MFPT [Fig.~\ref{fig:mfpt_pe_gamma}(b)]. For selected cases in the small-Pe regime, the corresponding FPT probability densities shown in Fig.~\ref{fig:mfpt_pe_gamma}(c) decay rapidly and show no distinct features for different~$\vartheta_0$.

\paragraph{Computer simulations.--} To corroborate our analytical results, we simulate the equations of motion [Eq.~\eqref{eq:langevin_theta}] employing the Euler-Maruyama scheme: 
\begin{subequations}
\begin{align}
\mathbf{r}(t+ \Delta t) &=  \mathbf{r}(t) + v \mathbf{e}(t) \Delta t + \sqrt{2D \Delta t} \mathbf{N}_{\boldsymbol\eta}, \\ 
\vartheta(t+ \Delta t) &= \vartheta(t) +\sqrt{2D_{\mathrm{rot}} \Delta t} N_{\xi},
\end{align}
\end{subequations}
where  $\mathbf{N}_{\boldsymbol\eta }(0, 1)$ and $N_{\xi}(0, 1)$ are independent standard normal random variables with zero mean and unit variance and the time step is set to $\Delta t = 10^{-8} \tau$ ($\Delta t = 10^{-4} \tau$) when studying absorbing (hard) walls. We simulate $2\times10^5$ trajectories to obtain reliable statistics.

\end{document}